\documentclass[aps,prb,twocolumn,amsmath,amssymb,groupedaddress,uplatex]{revtex4-2}

\usepackage{graphicx}
\usepackage{dcolumn}
\usepackage{bm}
\usepackage{multirow}
\usepackage{braket}
\usepackage{color}
\usepackage{ulem}
\usepackage{booktabs}

\begin{document}
\title{
Skyrmion crystal and spiral phases in centrosymmetric bilayer magnets\\ with staggered Dzyaloshinskii-Moriya interaction
}
\author{Satoru~Hayami}
\affiliation{Department of Applied Physics, the University of Tokyo, Tokyo 113-8656, Japan}

\begin{abstract}
We theoretically study a stabilization mechanism of the skyrmion crystal in centrosymmetric magnets with a bilayer structure. 
We show that the interplay between a layer-dependent staggered Dzyaloshinskii-Moriya interaction that arises from the absence of local inversion symmetry and the interlayer exchange interaction gives rise to a plethora of multiple-$Q$ states including the skyrmion crystal with a quantized topological number. 
By performing the simulated annealing for the bilayer triangular-lattice model under an external magnetic field, we demonstrate that the skyrmion forms the triangular-shaped crystals with different helicities in each layer owing to the staggered Dzyaloshinskii-Moriya interaction. 
Although the relative positions of the skyrmion core in each layer are different depending on the sign of the interlayer exchange interactions, the skyrmion crystal phases robustly appear under both ferromagnetic and antiferromagnetic interlayer interactions. 
We also find another two triple-$Q$ states with a uniform scalar chirality but without a quantized topological number in the low- and high-field regions. 
Especially, the low-field triple-$Q$ state exhibits the opposite sign of the scalar chirality to the skyrmion crystal, which are not found in the single-layer system. 
Our results indicate that the lack of local inversion symmetry in the lattice structure is another key ingredient to induce topological spin textures in centrosymmetric magnets. 
\end{abstract}

\maketitle

\section{Introduction}

The emergence of nontrivial topological spin textures has drawn considerable interest in condensed matter physics. 
Among them, a magnetic skyrmion characterized by a swirling spin texture with an integer of a topological number has been extensively studied in recent years, as it gives rise to intriguing physical phenomena caused by entanglement between topology and magnetism, such as the topological Hall effect~\cite{Bogdanov89,Bogdanov94,rossler2006spontaneous,nagaosa2013topological}. 
Since the discovery of the crystal formation of the magnetic skyrmion in the cubic chiral magnet MnSi~\cite{Muhlbauer_2009skyrmion}, known as the magnetic skyrmion crystal (SkX), it has been found in various noncentrosymmetric magnets~\cite{Tokura_doi:10.1021/acs.chemrev.0c00297}, such as another chiral magnets~\cite{yu2010real,yu2011near,seki2012observation,Adams2012,Seki_PhysRevB.85.220406,tokunaga2015new,karube2016robust,Li_PhysRevB.93.060409}, the polar magnets~\cite{heinze2011spontaneous,kezsmarki_neel-type_2015,Kurumaji_PhysRevLett.119.237201}, and the other noncentrosymmetric magnets~\cite{nayak2017discovery,peng2020controlled}. 
In these materials, the Dzyaloshinskii-Moriya (DM) interaction, which arises from relativistic spin-orbit coupling without inversion symmetry~\cite{dzyaloshinsky1958thermodynamic,moriya1960anisotropic}, is an important ingredient to realize the SkX. 
More specifically, an essence to stabilize the SkX is the competition between the ferromagnetic (FM) exchange interaction and the DM interaction under an external magnetic field~\cite{rossler2006spontaneous,Yi_PhysRevB.80.054416,Binz_PhysRevLett.96.207202,Binz_PhysRevB.74.214408}. 
Moreover, an unconventional short-period SkX and a hedgehog lattice with a three-dimensional topological spin texture have been observed in the chiral magnets EuPtSi~\cite{kakihana2018giant,kaneko2019unique,kakihana2019unique,tabata2019magnetic} and MnSi$_{1-x}$Ge$_{x}$, respectively~\cite{Binz_PhysRevB.74.214408,Park_PhysRevB.83.184406,Yang2016,tanigaki2015real,kanazawa2017noncentrosymmetric,fujishiro2019topological}. 
Although the above stabilization mechanism is not directly applied to the short-period SkX and hedgehog lattice, several theoretical studies have shown that the interplay between the DM interaction and the multiple-spin interactions can give rise to such short-period topological objects~\cite{heinze2011spontaneous,Hayami_PhysRevLett.121.137202,brinker2019chiral,Okumura_PhysRevB.101.144416,Mankovsky_PhysRevB.101.174401,paul2020role,Brinker_PhysRevResearch.2.033240,lounis2020multiple,grytsiuk2020topological,Kathyat_PhysRevB.103.035111,hayami2021field,Mendive-Tapia_PhysRevB.103.024410,hayami2021topological,Hayami_PhysRevB.104.094425,kato2021spin}.

On the other hand, the SkX and other topological spin textures have been recently observed in centrosymmetric lattice structures, such as the hexagonal magnets Gd$_2$PdSi$_3$~\cite{Saha_PhysRevB.60.12162,kurumaji2019skyrmion,sampathkumaran2019report,Hirschberger_PhysRevB.101.220401,Kumar_PhysRevB.101.144440,Spachmann_PhysRevB.103.184424} and  Gd$_3$Ru$_4$Al$_{12}$~\cite{hirschberger2019skyrmion,Hirschberger_10.1088/1367-2630/abdef9}, the tetragonal magnet GdRu$_2$Si$_2$~\cite{khanh2020nanometric,Yasui2020imaging}, and the cubic magnet SrFeO$_3$~\cite{Ishiwata_PhysRevB.84.054427,Ishiwata_PhysRevB.101.134406,Rogge_PhysRevMaterials.3.084404,Onose_PhysRevMaterials.4.114420}. 
From the theoretical point of view, these nontrivial topological spin textures in centrosymmetric magnets are brought about by the frustrated exchange interaction~\cite{Okubo_PhysRevLett.108.017206,leonov2015multiply,Lin_PhysRevB.93.064430,Hayami_PhysRevB.93.184413,batista2016frustration,Utesov_PhysRevB.103.064414,Wang_PhysRevB.103.104408}, the Ruderman-Kittel-Kasuya-Yosida interaction with/without magnetic anisotropy~\cite{Wang_PhysRevLett.124.207201,yambe2021skyrmion,mitsumoto2021replica}, and the multiple-spin interactions~\cite{Akagi_PhysRevLett.108.096401,Ozawa_PhysRevLett.118.147205,Hayami_PhysRevB.95.224424,Hayami_PhysRevB.99.094420,Simon_PhysRevMaterials.4.084408,hayami2020multiple,Hayami_PhysRevB.103.024439,Hayami_PhysRevB.103.054422,hayami2021topological,Eto_PhysRevB.104.104425,Hayami_10.1088/1367-2630/ac3683,hayami2021phase}. 
These studies open up a possibility to the realization of the SkX even in the centrosymmetric lattice systems, which provides still active research fields in both theory and experiment.

In the present study, we discuss another intriguing mechanism of the SkX in centrosymmetric magnets with the DM interaction. 
We consider the proper lattice structure where the inversion symmetry is preserved globally but broken intrinsically at atomic sites dubbed the local inversion symmetry breaking~\cite{zhang2014hidden,Hayami_PhysRevB.90.024432,Fu_PhysRevLett.115.026401,Razzoli_PhysRevLett.118.086402,hayami2016emergent,gotlieb2018revealing,Huang_PhysRevB.102.085205,Ishizuka_PhysRevB.98.224510}.
In this situation, the effect of the DM interaction appears to be canceled out due to the presence of global inversion symmetry, but it still remains in a sublattice-dependent form. 
For example, the zigzag~\cite{Yanase_JPSJ.83.014703,Hayami_doi:10.7566/JPSJ.84.064717,Hayami_doi:10.7566/JPSJ.85.053705,Sumita_PhysRevB.93.224507,cysne2021orbital}, honeycomb~\cite{Kane_PhysRevLett.95.226801,Hayami_PhysRevB.90.081115,yanagi2017optical,Yanagi_PhysRevB.97.020404}, and diamond~\cite{Fu_PhysRevLett.98.106803,Hayami_PhysRevB.97.024414,Ishitobi_doi:10.7566/JPSJ.88.063708} structures are typical prototypes with the sublattice-dependent DM interaction. 
Another example is a bilayer structure system where the sign of the DM interaction is opposite for the different two layers~\cite{hitomi2014electric,hitomi2016electric,yatsushiro2020odd,Yatsushiro_PhysRevB.102.195147}. 
Although the above lattice structures with the sublattice degree of freedom have attracted great interest owing to the findings of an antiferromagnetic (AFM) SkX~\cite{Rosales_PhysRevB.92.214439,zhang2016antiferromagnetic,Gobel_PhysRevB.96.060406,Kravchuk_PhysRevB.99.184429,gao2020fractional,Tome_PhysRevB.103.L020403} and intriguing dynamics~\cite{zhang2016magnetic,Zhang_PhysRevB.94.064406,koshibae2017theory,hrabec2017current,Shen_PhysRevB.98.134448,ang2019bilayer,Xia_PhysRevApplied.11.044046}, the effect of the sublattice-dependent DM interaction on the stabilization of the SkX has not been fully clarified yet~\cite{Diaz_hysRevLett.122.187203,Fang_PhysRevMaterials.5.054401}. 

Motivated by these studies, we here investigate the SkX formation in the centrosymmetric layered system by focusing on the role of the layer-dependent staggered DM interaction on the stabilization of the spiral and multiple-$Q$ states. 
Specifically, we consider the bilayer triangular-lattice system, where the layers are coupled by the FM or AFM interlayer interaction. 
We construct a magnetic phase diagram against the interlayer interaction and the magnetic field, and obtain eight phases including the SkX by performing the simulated annealing.  
We show that the SkXs with different helicities are stabilized on each layer as a result from the opposite sign of the DM interaction.  
We find that the spin textures of the SkX are clearly different between the FM- and AFM-coupled bilayers, although the same topological charge is obtained in both cases: 
The core positions of the SkXs on the different layers are different (the same) for the FM(AFM)-stacked cases. 
Furthermore, we find that two types of triple-$Q$ states with the uniform scalar chirality appear while increasing and decreasing the magnetic field in the SkX phase. 
In particular, the low-field triple-$Q$ state with the opposite sign of the scalar chirality to the SkX is stabilized by the synergy between the interlayer coupling and the staggered DM interaction that are not obtained in the single-layer model. 
The results indicate that the layer degree of freedom with the staggered DM interaction might be another prototypes to realize various topological spin textures even in centrosymmetric lattice structures. 
We discuss the details of the obtained spin and chirality configurations in real and momentum spaces in each phase. 

The remainder of this paper is structured as follows. 
In Sec.~\ref{sec:Model and method}, we introduce the bilayer system consisting of two triangular-lattice planes with the staggered DM interaction. 
We also outline the simulated annealing. 
We discuss the numerical results in Sec.~\ref{sec:Results}. 
After presenting the magnetic phase diagram while changing the interlayer exchange coupling and the external magnetic field, we show the spin and chirality textures in the obtained phases. 
Section~\ref{sec:Summary} is devoted to a summary. 
In Appendix~\ref{sec:Spin configurations and structure factors of the other phases}, we show the real-space spin configurations and the spin and chirality structure factors in the magnetic phases that are not mentioned in the main text. 
We show the results for the different DM interactions in Appendix~\ref{sec:Results for different D}.

\section{Model and method}
\label{sec:Model and method}

\begin{figure}[htb!]
\begin{center}
\includegraphics[width=1.0 \hsize]{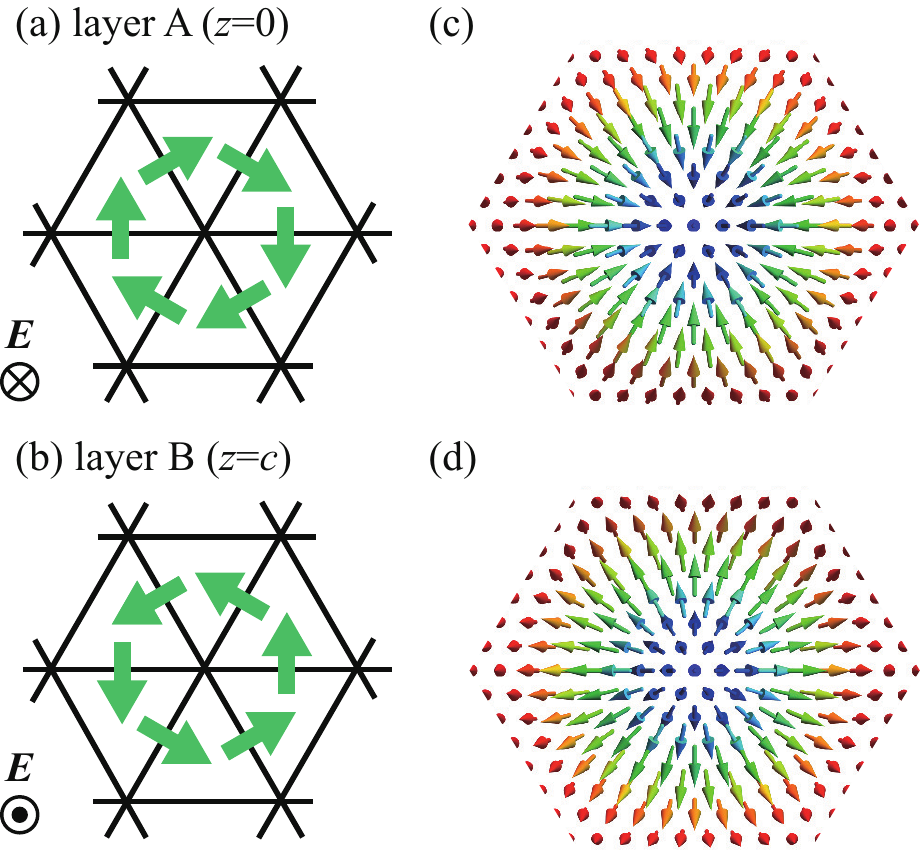} 
\caption{
\label{Fig:lattice}
(a), (b) The bilayer triangular-lattice system consisting of (a) the layer A and (b) the layer B under the local crystalline electric field $\bm{E}$ along the opposite directions. 
The green arrows represent the DM vectors. 
(c), (d) The N\'eel-type skyrmion spin textures with the opposite helicities, which are related to the DM vectors in (a) and (b). 
}
\end{center}
\end{figure}

We consider a bilayer system consisting of two triangular-lattice planes. 
We take the triangular planes in the $xy$ plane and the stacking direction along the $z$ direction; we label the lower and upper layers as layer A and layer B, which are separated by $c=1$. 
Considering the bilayer structure, inversion center is not located at the lattice sites but at the bond center between the layers. 
Thus, there is a local crystalline electric field on each layer, whose directions are opposite to each other. 
In such a situation, the layer-dependent staggered DM interaction appears so as to satisfy global inversion symmetry in the system; the DM vectors are lied perpendicular to the intralayer bond direction and the $z$ direction in the opposite way for the different layers, as shown in Figs.~\ref{Fig:lattice}(a) and \ref{Fig:lattice}(b). 

The bilayer triangular spin model incorporating the effect of the staggered DM interaction is given by 
\begin{align}
\label{eq:Ham}
\mathcal{H}=&\sum_{\gamma}\mathcal{H}^{\perp}_{\gamma}+\mathcal{H}^{\parallel}+\mathcal{H}^{{\rm Z}}, \\
\label{eq:Ham_perp}
\mathcal{H}^{\perp}_{\gamma}=&  \sum_{ij} \left[ J_{ij} \bm{S}_{i} \cdot \bm{S}_{j}-    \bm{D}_{ij}^{(\gamma)} \cdot  (\bm{S}_{i} \times \bm{S}_{j}) \right], \\
\label{eq:Ham_parallel}
\mathcal{H}^{\parallel}=& J_{\parallel} \sum_{i} \bm{S}_i \cdot \bm{S}_{i+\hat{z}},\\
\label{eq:Ham_Zeeman}
\mathcal{H}^{{\rm Z}}=&-H \sum_i S_i^z. 
\end{align}
The total Hamiltonian $\mathcal{H}$ consists of three parts: the intralayer Hamiltonian $\mathcal{H}^{\perp}_{\gamma}$ ($\gamma={\rm A}, {\rm B}$ is the layer index), the interlayer Hamiltonian $\mathcal{H}^{\parallel}$, and the Zeeman Hamiltonian $\mathcal{H}^{{\rm Z}}$. 
The first term of $\mathcal{H}^{\perp}_{\gamma}$ represents the layer-independent exchange interaction $J_{ij}$ and the second term represents the layer-dependent staggered DM interaction, i.e., $\bm{D}_{ij}^{({\rm A})}=-\bm{D}_{ij}^{({\rm B})}$ with the same magnitude $|\bm{D}_{ij}^{({\rm A})}|=|\bm{D}_{ij}^{({\rm B})}|=D_{ij}$. 
We take the directions of the staggered DM vector along the directions shown in Figs.~\ref{Fig:lattice}(a) and \ref{Fig:lattice}(b). 
The two layers are coupled via the exchange coupling $J_{\parallel}$ in Eq.~(\ref{eq:Ham_parallel}). 
The effect of an external magnetic field is introduced by the Zeeman coupling with the field strength $H$ along the $z$ direction in Eq.~(\ref{eq:Ham_Zeeman}).
In the model in Eq.~(\ref{eq:Ham}), we neglect long-range dipole-dipole interactions for simplicity. 

The magnetic phases while changing $H$ were investigated for the model with the FM exchange interaction and the DM interaction between the nearest-neighbor spins in the absence of $J_{\parallel}$, i.e. the single-layer model, ~\cite{Yi_PhysRevB.80.054416,Mochizuki_PhysRevLett.108.017601,Rowland_PhysRevB.93.020404}:  
the single-$Q$ cycloidal spiral state for the low-field region, the N\'eel SkX for the intermediate-field region, and the fully-polarized state for the high-field region. 
The N\'eel SkX is described by a superposition of three cycloidal spirals connected by threefold rotational symmetry of the triangular-lattice structure. 
The spin helicity in the cycloidal spiral state and the SkX is determined by the sign of the DM interaction. 
We show the skyrmion spin textures stabilized on the layers A and B in Figs.~\ref{Fig:lattice}(c) and \ref{Fig:lattice}(d), respectively, where the inplane spin directions around the skyrmion core are opposite to each other. 

Based on the magnetic instabilities in the single-layer system, we here focus on the effect of the interlayer exchange coupling $J_{\parallel}$ on the stabilization of the single-$Q$ spiral and the N\'eel SkX. 
Owing to the opposite helicity in the spiral spin textures for the different layers, a magnetic frustration occurs irrespective of the FM and AFM interlayer interactions; the FM (AFM) interaction leads to an energy cost in the $xy$($z$)-spin component when the core positions of the SkXs in both layers are the same, as found in Figs.~\ref{Fig:lattice}(c) and \ref{Fig:lattice}(d). 
Furthermore, the model Hamiltonian possesses spatial inversion symmetry when considering the bilayer structure with the inversion center at bonds between the layers, which makes optimized spin configurations nontrivial. 

In order to examine such effects of the interlayer exchange coupling and the staggered DM interaction, we simplify the intralayer Hamiltonian $\mathcal{H}^{\perp}_1+\mathcal{H}^{\perp}_2$ as 
\begin{align}
\label{eq:Ham_perp2}
\tilde{\mathcal{H}}^{\perp}=&  \sum_{\nu} \sum_{\gamma} \Big[ -J \bm{S}^{(\gamma)}_{\bm{Q}_{\nu}} \cdot \bm{S}^{(\gamma)}_{-\bm{Q}_{\nu}}- i   \bm{D}^{(\gamma)}_\nu \cdot ( \bm{S}^{(\gamma)}_{\bm{Q}_{\nu}} \times \bm{S}^{(\gamma)}_{-\bm{Q}_{\nu}}) \Big],  
\end{align}
where $\bm{S}^{(\gamma)}_{\bm{Q}_{\nu}}$ is the Fourier transform of $\bm{S}_i$ with wave vector $\bm{Q}_\nu$ for the layer $\gamma$. 
In Eq.~(\ref{eq:Ham_perp2}), we extract the dominant $\bm{q}$ contributions from $\sum_{\bm{q}} [J^{(\gamma)}_{\bm{q}} \bm{S}^{(\gamma)}_{\bm{q}} \cdot \bm{S}^{(\gamma)}_{-\bm{q}}+i   \bm{D}^{(\gamma)}_{\bm{q}} \cdot ( \bm{S}^{(\gamma)}_{\bm{q}} \times \bm{S}^{(\gamma)}_{-\bm{q}})]$, which corresponds to the Fourier transform of $\mathcal{H}^{\perp}_{\gamma}$ in Eq.~(\ref{eq:Ham_perp}), by supposing six global energy minima in momentum space so as to satisfy the rotational symmetry of the bilayer triangular lattice~\cite{leonov2015multiply,Hayami_PhysRevB.103.224418}. 
We suppose global minima at $\bm{Q}_1=(\pi/3,0)$, $\bm{Q}_2=(-\pi/6,\sqrt{3}\pi/6)$, $\bm{Q}_3=(-\pi/6,-\sqrt{3}\pi/6)$, $\bm{Q}_4=-\bm{Q}_1$, $\bm{Q}_5=-\bm{Q}_2$, and $\bm{Q}_6=-\bm{Q}_3$; $J \equiv J^{(\gamma)}_{\bm{Q}_\nu}$ and $\bm{D}^{(\gamma)}_\nu \equiv \bm{D}^{(\gamma)}_{\bm{Q}_\nu}$. 
We neglect the contributions from the other $\bm{q}$ components in the interactions for simplicity. 
Hereafter, we fix $J=1$ and $|\bm{D}^{(\gamma)}_{\bm{Q}_\nu}|=D=0.2$ and take $J_{\parallel}$ and $H$ as the parameters. 
The model in Eq.~(\ref{eq:Ham_perp2}) in the absence of $\mathcal{H}^{\parallel}$ reproduces the single-$Q$ cycloidal spiral and the SkX while changing the magnetic field, as the model in Eq.~(\ref{eq:Ham_perp}) does, which will be shown later.

In the following, we perform the simulated annealing to determine the magnetic phase diagram of the model $\tilde{\mathcal{H}}=\tilde{\mathcal{H}}^{\perp}+\mathcal{H}^{\parallel}+\mathcal{H}^{{\rm Z}}$ on the bilayer triangular lattice.  
In the simulations, we gradually reduce the temperature from high temperature with a rate $T_{n+1}=\alpha T_n$, where $T_n$ is the $n$th-step temperature ($T_0=0.1$-$1.0$) and $\alpha=0.999995$. 
The final temperature is set as $T=0.001$. 
After reaching the final temperature, we perform $10^5$-$10^6$ Monte Carlo sweeps for measurements. 
The update of the spin configuration is performed based on the standard Metropolis local updates. 
The total number of spins are taken as $N=2\times 96^2$. 

The magnetic phases are identified by the spin and chirality configurations at the lowest temperature. 
The spin structure factor is represented by 
\begin{align}
S_{\eta}^\alpha(\bm{q})= \frac{1}{N} \sum_{j,l \in \eta} S^{\alpha}_j S^{\alpha}_l e^{i\bm{q}\cdot (\bm{r}_j-\bm{r}_l)}, 
\end{align}
for $\alpha=x,y,z$. 
The site indices $j$ and $l$ are taken for the layer $\eta=1,2$. 
The total spin structure factor in the system is $S_{s}^\alpha(\bm{q})=S_{{\rm A}}^\alpha(\bm{q})+S_{{\rm B}}^\alpha(\bm{q})$. 
We also compute $S_{\eta}^{xy}(\bm{q})=S_{\eta}^x(\bm{q})+S_{\eta}^y(\bm{q})$.
The net magnetization for each layer is given by $M^\alpha_{\eta}=(1/N)\sum_{i \in \eta}S^{\alpha}_{i}$. 

The spin scalar chirality is represented by 
\begin{align}
\chi^{\rm sc}_{\eta} = \frac{1}{N} \sum_{\bm{R}\in \eta} \bm{S}_{j} \cdot (\bm{S}_k \times \bm{S}_l),
\end{align}
where $\bm{R}$ represents the position vector at the centers of triangles; the sites $j$, $k$, and $l$ form the triangle at $\bm{R}$ in the counterclockwise order. 
The local chirality is represented by $\chi_{\bm{R}}=\bm{S}_{j} \cdot (\bm{S}_k \times \bm{S}_l)$. 
The magnetic ordering with nonzero $\chi^{\rm sc}=\chi^{\rm sc}_{\rm A}+ \chi^{\rm sc}_{\rm B}$ exhibits the topological Hall effect. 
We also calculate the scalar chirality structure factor is given by 
\begin{align}
\label{eq:chiralstructurefactor}
S^{\chi}_{\eta}(\bm{q})= \frac{1}{N}\sum_{\mu}\sum_{\bm{R},\bm{R}' \in \mu}  \chi_{\bm{R}}
\chi_{\bm{R}'} e^{i \bm{q}\cdot (\bm{R}-\bm{R}')}, 
\end{align}
where $\mu=(u, d)$ represent upward and downward triangles, respectively.

\section{Results}
\label{sec:Results}

\begin{figure}[htb!]
\begin{center}
\includegraphics[width=1.0 \hsize]{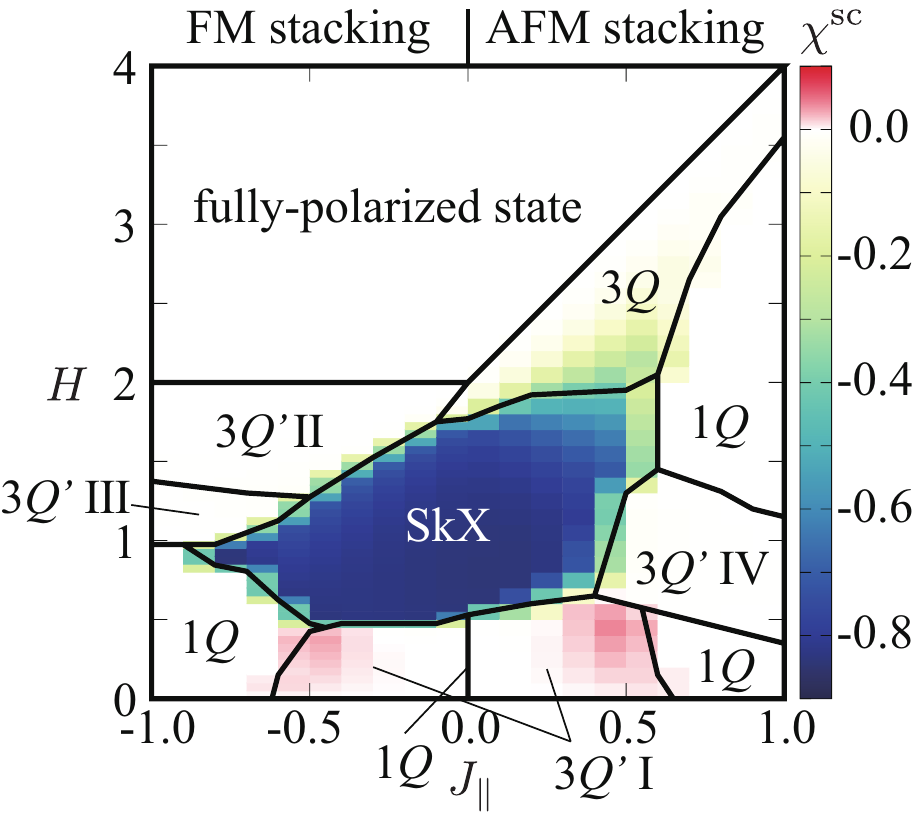} 
\caption{
\label{Fig:PD}
The magnetic phase diagram while changing the interlayer exchange interaction $J_{\parallel}$ and the external magnetic field $H$. 
The color plot represents the spin scalar chirality $\chi^{\rm sc}$. 
The regions for $J_{\parallel}>0$ and $J_{\parallel}<0$ represent the cases of the  antiferromagnetic (AFM) and ferromagnetic (FM) stackings, respectively. 
}
\end{center}
\end{figure}

\begin{figure}[htb!]
\begin{center}
\includegraphics[width=1.0 \hsize]{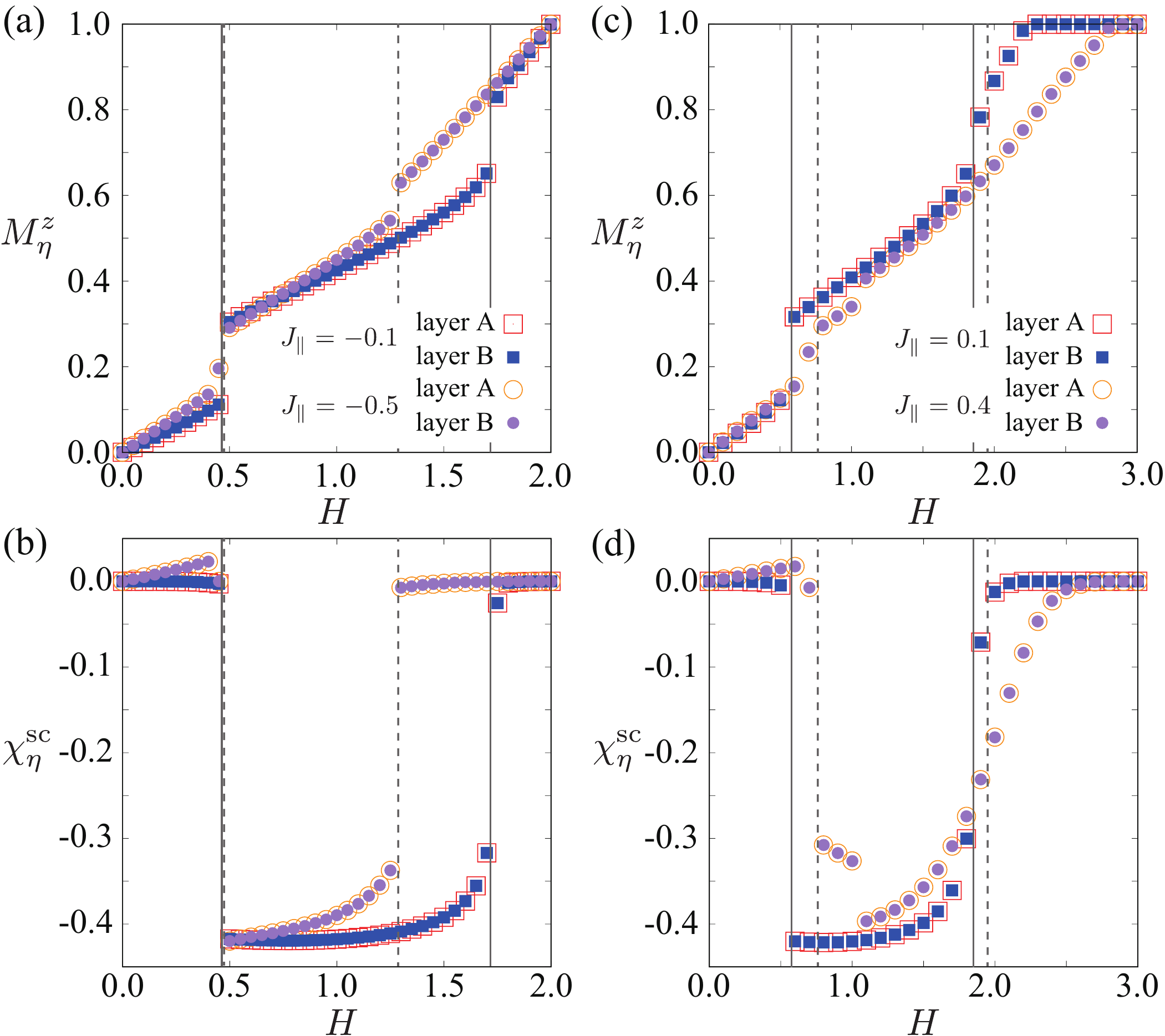} 
\caption{
\label{Fig:Mag_chirality}
$H$ dependence of (a), (c) the magnetization $M^z_\eta$ and (b), (d) the scalar chirality $\chi^{\rm sc}_\eta$ for the layers $\eta=$A and B at (a), (b) $J_{\parallel}=-0.1$ and $-0.5$ and (c), (d) $J_{\parallel}=0.1$ and $0.4$. 
The solid (dashed) vertical lines represent the phase transitions between the SkX and the other magnetic states for the small (large) values of $|J_{\parallel}|$. 
}
\end{center}
\end{figure}

Figure~\ref{Fig:PD} shows the magnetic phase diagram obtained by simulated annealing down to $T=0.001$ while changing the interlayer exchange coupling $J_{\parallel}$ and the magnetic field $H$ on the bilayer triangular lattice. 
When $J_{\parallel}=0$, the present system reduces to the single-layer system. 
The single-$Q$ spiral (1$Q$) state is stabilized at zero field, which turns into the SkX, the 3$Q$ state, and the fully-polarized state while increasing $H$. 
Such a tendency of the transitions is similar to that for the model with the ferromagnetic and the DM interaction in the single-layer system~\cite{Yi_PhysRevB.80.054416,Mochizuki_PhysRevLett.108.017601}. 

The introduction of $J_{\parallel}$ changes the magnetic phase diagram.
The positive (negative) $J_{\parallel}$ represents the AFM(FM)-stacked case. 
As shown in Fig.~\ref{Fig:PD}, we obtain seven magnetic states including the SkX except for the fully-polarized state with $\bm{S}_i = (0,0,1)$ in the high-field region. 
Among the magnetic states, the SkX, 1$Q$, and triple-$Q$ spiral (3$Q'$ I) are robustly stabilized in both FM- and AFM-stacked regions, while the triple-$Q$ states denoted as 3$Q$ and 3$Q'$ IV (3$Q'$ II and 3$Q'$ III) appears in the AFM(FM)-stacked region, where $Q'$ represents the different amplitudes of the constituent waves. 
Each magnetic state is characterized by the different spin and chirality configurations, as shown below. 
In addition, we find that three out of seven states exhibit the uniform spin scalar chirality, $\chi_{\rm sc}$: negative chirality in the SkX and the 3$Q$ state and the positive one in the 3$Q'$ I state, as shown in the color plot of $\chi_{\rm sc}$ in Fig~\ref{Fig:PD}. 
Especially, only the SkX phase has the quantized skyrmion number $-1$ in each layer in the magnetic unit cell.  
$\chi_{\rm sc}$ in the other phases is negligibly small.

We show the $H$ dependence of the magnetization $M^z_{\eta}$ and the scalar chirality $\chi^{\rm sc}_{\eta}$ for each layer in the FM stacking in Figs.~\ref{Fig:Mag_chirality}(a) and \ref{Fig:Mag_chirality}(b) and the AFM stacking in Figs.~\ref{Fig:Mag_chirality}(c) and \ref{Fig:Mag_chirality}(d). 
The parameters of $J_{\parallel}$ are taken for those where the SkX is stabilized in the intermediate-field region. 
As shown in Fig.~\ref{Fig:Mag_chirality}, there are no differences of $M^z_{\eta}$ and $\chi_{\eta}^{\rm sc}$ between the layers A and B in spite of the opposite directions of the DM vectors. 
Furthermore, one finds that the SkX and the 3$Q$ state exhibit a negative scalar chirality, while the 3$Q'$ I state exhibits a positive one. 
The phase transitions between the SkX and the other magnetic states are of first order with jumps of $M^z_{\eta}$ and $\chi_{\eta}^{\rm sc}$. 
The other transitions in Fig.~\ref{Fig:PD} are as follows: the phase transitions between the 1$Q$ and 3$Q'$ I states for nonzero $J_{\parallel}$, between the $3Q'$ I and 3$Q'$ IV states, and between the 1$Q$ and 3$Q$ are of first order, while the others are of second order. 

\begin{figure*}[htb!]
\begin{center}
\includegraphics[width=1.0 \hsize]{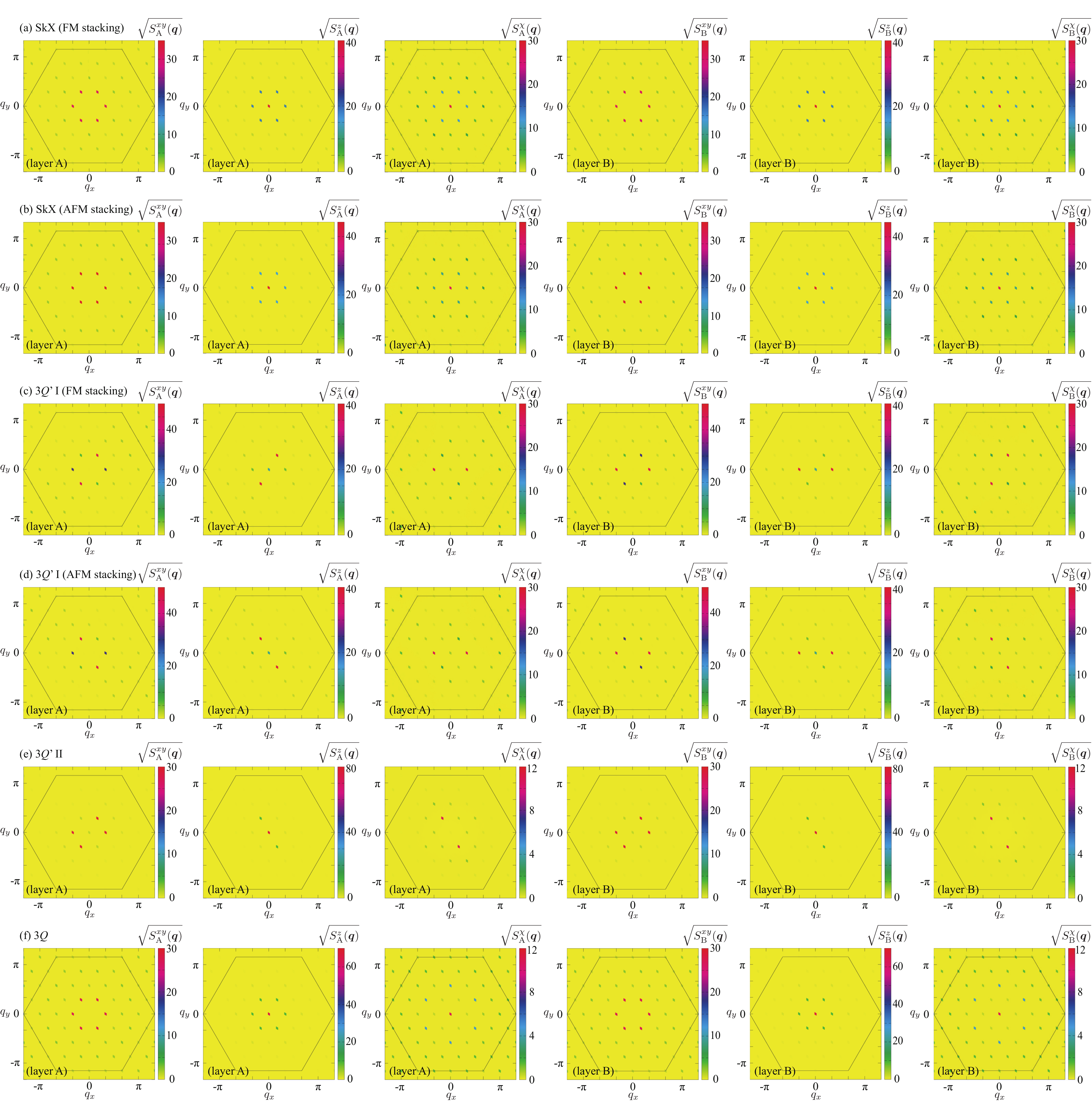} 
\caption{
\label{Fig:SkX_sq}
(Left and second left) The square root of the $xy$ and $z$ components of the spin structure factor for the layer A in (a) the SkX at $J_{\parallel}=-0.5$ and $H=0.85$, (b) the SkX at $J_{\parallel}=0.4$ and $H=1.3$, (c) the 3$Q'$ I state at $J_{\parallel}=-0.5$ and $H=0.4$, (d) the 3$Q'$ I state at $J_{\parallel}=0.5$ and $H=0.5$, (e) the 3$Q'$ II state at $J_{\parallel}=-1$ and $H=1.6$, and (f) the 3$Q$ state at $J_{\parallel}=0.5$ and $H=2.1$.
Black hexagons represent the first Brillouin zone.   
(Middle left) The square root of the chirality structure factor for the layer A. 
The right three panels represent the data for the layer B corresponding to the left three ones. 
}
\end{center}
\end{figure*}

In the following, we discuss the detailed spin and chirality configurations of the SkX in Sec.~\ref{sec:Skyrmion crystal}, the 3$Q'$ I state in Sec.~\ref{sec:Low-field triple-$Q$ state}, and the 3$Q'$ II and 3$Q$ states in Sec.~\ref{sec:High-field triple-$Q$ states}. 
The momentum-space spin and chirality structure factors are shown in Fig.~\ref{Fig:SkX_sq} and the snapshots in terms of the real-space spin and chirality obtained by simulated annealing in Figs.~\ref{Fig:SkX}, \ref{Fig:3QI}, and \ref{Fig:highfield}. 
In both real and momentum spaces, we show the spin- and chirality-related quantities in each layer. 
In addition, we show the averaged spin and chirality configurations over the layers in order to clearly show the similarity and difference between the layers A and B. 
The spin and chirality structures in the other magnetic states, $1Q$, 3$Q'$ III, and 3$Q'$ IV, are discussed in Appendix~\ref{sec:Spin configurations and structure factors of the other phases}.

\subsection{Skyrmion crystal}
\label{sec:Skyrmion crystal}

\begin{figure*}[htb!]
\begin{center}
\includegraphics[width=1.0 \hsize]{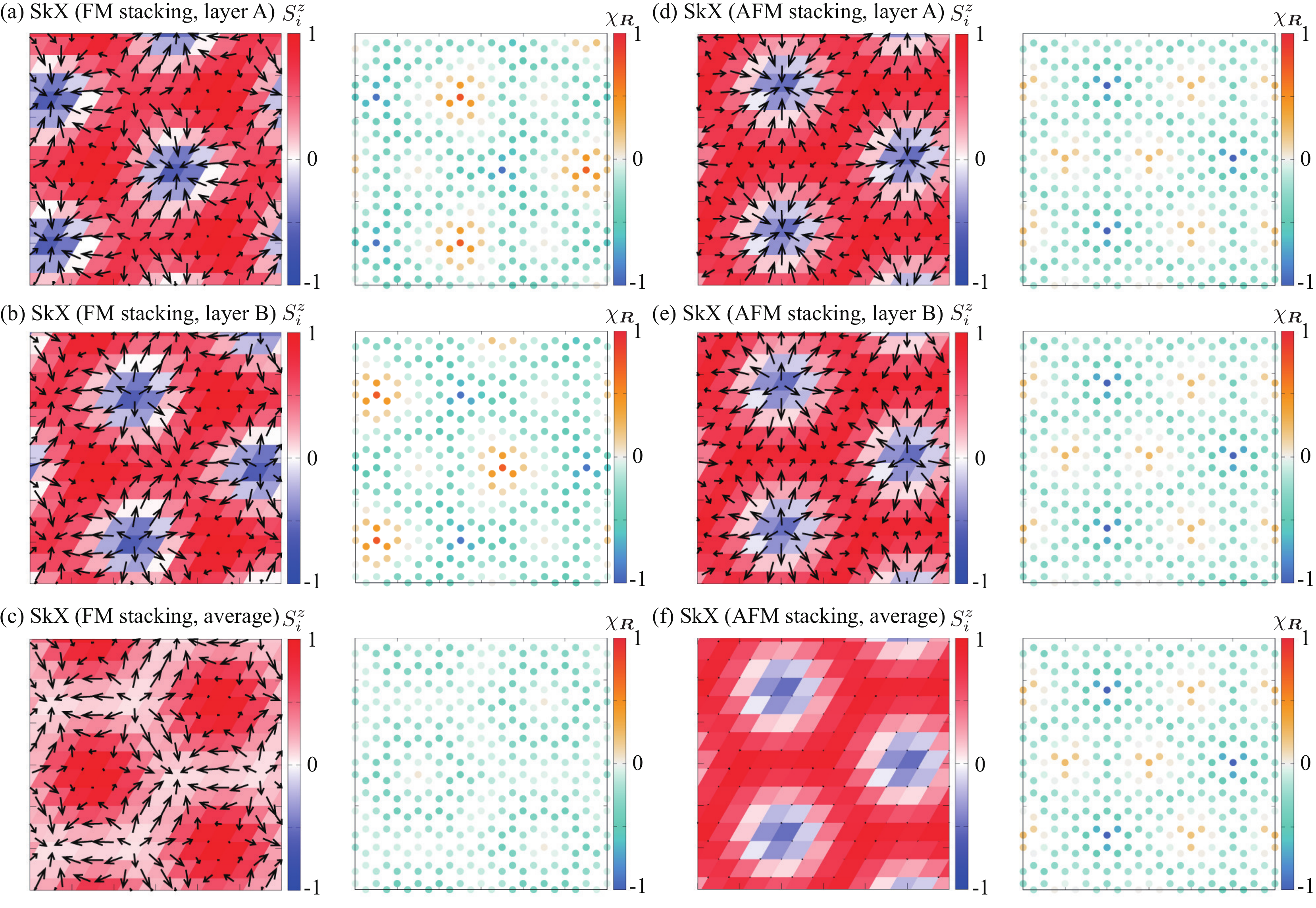} 
\caption{
\label{Fig:SkX}
(Left) Real-space spin configurations of the SkXs on (a), (d) the layer A and (b), (e) the layer B in (a), (b) the FM stacked case at $J_{\parallel}=-0.5$ and $H=0.85$ and (d), (e) the AFM stacked case at $J_{\parallel}=0.4$ and $H=1.3$. 
(c), (f) The averaged spin configurations for the layers A and B in (c) the FM stacked case and (f) the AFM stacked case. 
The arrows represent the $xy$ components of the spin moment and the color shows the $z$ component. 
(Right) Real-space scalar chirality configurations calculated at each triangle plaquette. 
}
\end{center}
\end{figure*}

We discuss the SkX stabilized in the intermediate-field region in Fig.~\ref{Fig:PD} in this section. 
The SkX robustly appears for both FM and AFM interlayer interactions. 
The origin of the SkX is obviously attributed to the staggered DM interaction inherent of the bilayer structure.  
As detailed below, we obtain the skyrmion spin textures with the opposite helicities in each layer, as inferred from the DM vector [see also Figs.~\ref{Fig:lattice}(c) and \ref{Fig:lattice}(d)]. 
The emergence of the SkX in the present bilayer system indicates the importance of the layer degree of freedom in the lattice structure, which gives a way to realize the SkX in centrosymmetric systems without relying on the frustrated exchange interactions and multiple-spin interactions. 

In each layer, the SkX is characterized by a superposition of three cycloidal spirals along the $\bm{Q}_1$, $\bm{Q}_2$, and $\bm{Q}_3$ directions, and hence, it exhibits triple-$Q$ peaks with equal intensity in the spin and chirality structure factors. 
The resultant spin textures in real space are represented by a periodic array of the skyrmion core with $S_i^z \simeq -1$, which forms the triangular lattice. 
The vorticity around the skyrmion core is fixed at $+1$, while the helicity depends on the sign of the DM interaction. 
In the present bilayer system, the helicities around the skyrmion core should be opposite between the layers A and B due to the staggered DM interaction. 
Indeed, we obtain such a tendency in the simulations in the parameter region where the SkX is stabilized irrespective of the FM and AFM interlayer interactions; the direction of the inplane spins around the skyrmion core is inward for the layer A [Figs.~\ref{Fig:SkX}(a) and \ref{Fig:SkX}(d)], while that is outward for the layer B [Figs.~\ref{Fig:SkX}(b) and \ref{Fig:SkX}(e)]. 
In addition, there is a uniform scalar chirality in both layers with the same sign, which gives rise to the quantized skyrmion number of $-1$. 
This real-space threefold-symmetric spin and chirality textures on both layers are consistent with the triple-$Q$ peaks in the spin and chirality structure factors in momentum space, as  shown in Figs.~\ref{Fig:SkX_sq}(a) and \ref{Fig:SkX_sq}(b). 
There is no difference of the spin and chirality structure factors between the layers A and B in both cases of the FM and AFM interactions as well as $M^z_{\eta}$ and $\chi_{\eta}^{\rm sc}$ in Fig.~\ref{Fig:Mag_chirality}.

Meanwhile, a clear difference between the FM and AFM interlayer interactions is found in local spin and chirality configurations in a real-space picture. 
The skyrmion cores lie at the different positions on the layers A and B under the FM stacking as shown in Figs.~\ref{Fig:SkX}(a) and \ref{Fig:SkX}(b), while those lie at the same positions under the AFM stacking as shown in Figs.~\ref{Fig:SkX}(d) and \ref{Fig:SkX}(e). 
By closely looking into the real-space spin configurations, one finds that the SkXs are stacked so that the inplane spins on the two layers are aligned (anti)parallel to each other in the FM (AFM) interaction. 
In other words, the SkXs are stacked so as to gain the exchange energy in terms of the $xy$ spin component rather than the $z$ spin component. 
This seems to be reasonable, since the $xy$ spin contribution is larger than the $z$ spin contribution in the $\bm{Q}_\nu$ component of the spin structure factor, e.g., $S^{xy}_{\rm A}(\bm{Q}_1)/S_{\rm A}^{z}(\bm{Q}_1) \simeq 1.34$ at $J_{\parallel}=0$ and $H=0.85$. 
Furthermore, there is a slight difference of the constituent vortices between the FM and AFM interactions: the SkX stabilized by the FM interaction consists of the vortices with vorticity $-2$ around $S_i^z \simeq +1$, another with vorticity $+1$ around $S^z_i \simeq +1$, and the other with vorticity $+1$ around $S_i^z \simeq -1$ in each layer, while the SkX by the AFM interaction does not have the vortices with vorticity $-2$. 
Reflecting the difference, the scalar chirality distribution looks threefold(sixfold)-symmetric in the FM(AFM)-coupled SkX, as shown in the right panels of Figs.~\ref{Fig:SkX}(a) and \ref{Fig:SkX}(b) [Figs.~\ref{Fig:SkX}(d) and \ref{Fig:SkX}(e)]~\cite{comment_sixfold}. 

The different skyrmion core positions between the FM and AFM interactions result in a difference of the averaged spin textures over the layers. 
Figure~\ref{Fig:SkX}(c) shows the averaged spin textures in the case of the FM interaction, where all the spins have the positive $z$ component. 
In the regions where the skyrmion cores lie on the layer A or B, the $z$-spin component becomes small owing to the cancellation between the negative contribution from the skyrmion core on the layer A (B) and the positive contribution from the vortex core with the same helicity but different chirality from the skyrmion core on the layer B (A). 
Meanwhile, the vortex cores with vorticity $-2$ on the layers A and B are located at the same position, which leads to the meron spin texture with vorticity $-2$, i.e., the skyrmion number $-1$.  
Thus, the averaged spin texture in the FM-coupled SkX is the same as that in the meron crystal with vorticity $-2$. 

On the other hand, in the case of the AFM-coupled SkX, the $xy$ spin components of the layers A and B are cancelled out and only the $z$ spin component remains. 
Thus, the averaged spin texture looks like the magnetic bubble crystal, where the cores with $S_i^z \simeq -1$ form the triangular lattice, as shown in Fig.~\ref{Fig:SkX}(f).
Nevertheless, it is noted that the averaged scalar chirality is not cancelled out, since the SkXs with different helicities induce the same sign of the scalar chirality. 
This means that the AFM-coupled SkX has the skyrmion number of $-1$, since both the layers A and B have the equal skyrmion number of $-1$ in each layer [see also the right panel of Figs.~\ref{Fig:SkX}(d)-\ref{Fig:SkX}(f)]. 

While further increasing $J_{\parallel}$, the SkX is replaced with the other 1$Q$ and multiple-$Q$ states: the 1$Q$, 3$Q'$ II, and 3$Q'$ III states for the FM interlayer interaction $J_{\parallel}<0$ and the 1$Q$ and 3$Q'$ IV state for the AFM interaction $J_{\parallel}>0$ depending on the magnitude of the magnetic field. 
The critical value of $J_{\parallel}$ to destabilize the SkX is larger for the FM interaction than the AFM one, both of which is smaller than the intralayer interaction $J$.  
Thus, the bilayer system coupled via the FM interaction might be a suitable system to show the SkX in the intermediate field. 
We also discuss the stability of the SkX by changing the DM interaction in Appendix~\ref{sec:Results for different D}, where the large staggered DM interaction tends to favor the SkX like the noncentrosymmetric system with the uniform DM interaction.

\subsection{Low-field triple-$Q$ state}
\label{sec:Low-field triple-$Q$ state}

\begin{figure*}[htb!]
\begin{center}
\includegraphics[width=1.0 \hsize]{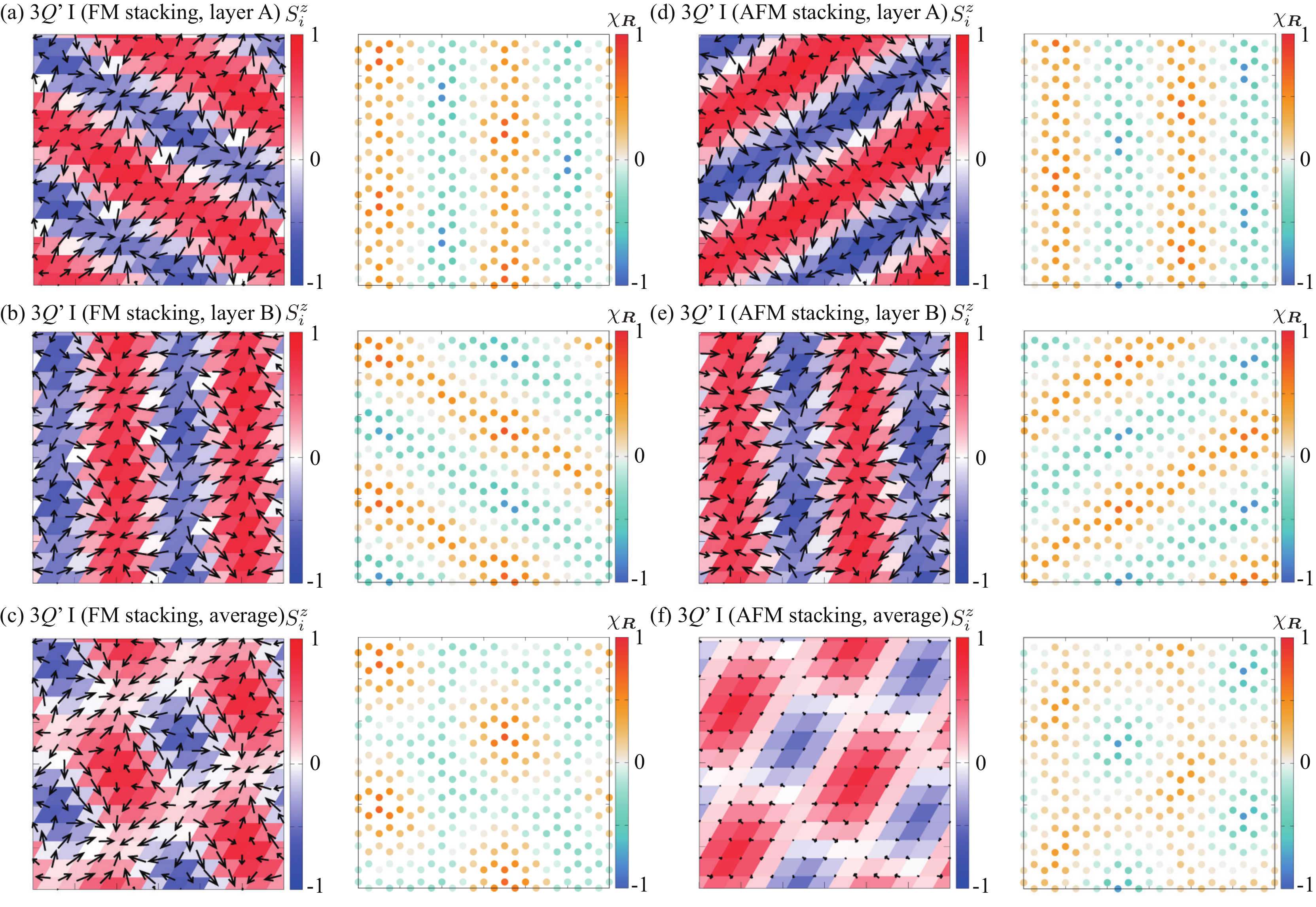} 
\caption{
\label{Fig:3QI}
(Left) Real-space spin configurations of the 3$Q'$ I state on (a), (d) the layer A and (b), (e) the layer B in (a), (b) the FM stacked case at $J_{\parallel}=-0.5$ and $H=0.4$ and (d), (e) the AFM stacked case at $J_{\parallel}=0.5$ and $H=0.5$. 
(c), (f) The averaged spin configurations for the layers A and B in (c) the FM stacked case and (f) the AFM stacked case. 
The arrows represent the $xy$ components of the spin moment and the color shows the $z$ component. 
(Right) Real-space scalar chirality configurations calculated at each triangle plaquette. 
}
\end{center}
\end{figure*}

We discuss the low-field magnetic phases, which are obtained while decreasing $H$ from the SkX phase. 
For $J_{\parallel}=0$, the 1$Q$ cycloidal spiral state is stabilized to gain the energy of the staggered DM interaction. 
The introduction of $J_{\parallel}$ leads to the multiple-$Q$ spin modulations from the 1$Q$ state. 
The resultant spin and chirality structure factors are similar between the FM and AFM stacked cases, as shown in Figs.~\ref{Fig:SkX_sq}(c) and \ref{Fig:SkX_sq}(d); in each layer, there are dominant single-$Q$ peaks at any of $(\bm{Q}_1, \bm{Q}_2, \bm{Q}_3)$ in both $xy$ and $z$ spin components, and  the subdominant double-$Q$ peaks with different intensities at the remaining two $\bm{Q}_\nu$. 
The smallest peak among $(\bm{Q}_1, \bm{Q}_2, \bm{Q}_3)$ vanishes at zero field; the zero-field state corresponds to the double-$Q$ state. 
The multiple-$Q$ spin textures in the presence of $J_{\parallel}$ are accompanied with the scalar chirality density waves at $\bm{Q}_1$-$\bm{Q}_3$. 
The dominant component of $S^{\chi}_{\eta}(\bm{q})$ is the second largest $\bm{Q}_\eta$ in the spin structure factor. 
This spin and chirality textures are similar to the chiral stripe state appearing in the itinerant electron model without the spin-orbit coupling~\cite{Solenov_PhysRevLett.108.096403,Ozawa_doi:10.7566/JPSJ.85.103703,yambe2020double}, but the origin is different with each other: The present multiple-$Q$ state is stabilized by the interplay between the staggered DM and interlayer interactions, while the chiral stripe state is stabilized by the multiple-spin interactions arising from the itinerant nature of electrons~\cite{Ozawa_doi:10.7566/JPSJ.85.103703}. 

There are two characteristic points in the low-field 3$Q'$ I state as a consequence of the bilayer system. 
The one is the layer-dependent $\bm{q}$-peak structure. 
As shown in Figs.~\ref{Fig:SkX_sq}(c) and \ref{Fig:SkX_sq}(d), the dominant $\bm{q}$ components are different for the layers A and B in both FM and AFM interactions. 
For example, in the case of the FM interaction in Fig.~\ref{Fig:SkX_sq}(c), the dominant peak in the spin structure factors $S^{xy}_{\rm A}(\bm{q})$ and $S^{z}_{\rm A}(\bm{q})$ lies at $\bm{Q}_3$, while that in $S^{xy}_{\rm B}(\bm{q})$ and $S^{z}_{\rm B}(\bm{q})$ lies at $\bm{Q}_1$. 
Accordingly, the dominant peak position in the scalar chirality structure factor $S^{\chi}_{\rm A}(\bm{q})$ is different from that in $S^{\chi}_{\rm B}(\bm{q})$. 
The difference is clearly found in the real-space spin and chirality configurations in Figs.~\ref{Fig:3QI}(a) and \ref{Fig:3QI}(b).  

The different dominant components in $S^{\alpha}_{\eta}(\bm{q})$ and $S^{\chi}_{\eta}(\bm{q})$ between the layers A and B are attributed to the staggered DM interaction, since it fixes the helicity of the spiral in an opposite way for the different layers. 
In such a situation, there is a frustration between the $xy$ and $z$ spin components similar to that in the SkX in Sec.~\ref{sec:Skyrmion crystal}. 
The present results indicate that the choice of the different $\bm{Q}_\nu$ in the layers A and B gains the energy of the layer-dependent staggered DM and interlayer interactions as much as possible. 
A similar situation happens in the case of the AFM interaction, as shown in Figs.~\ref{Fig:3QI}(d) and \ref{Fig:3QI}(e).

The other characteristic point is the nonzero uniform scalar chirality $\chi^{\rm sc}$, as shown in Fig.~\ref{Fig:PD}. 
Similar to the SkX, both layers take the same value of $\chi^{\rm sc}_\eta$, as shown in Figs.~\ref{Fig:Mag_chirality}(b) and \ref{Fig:Mag_chirality}(d).  
On the other hand, the chirality takes a positive value and the skyrmion number is not quantized in contrast to the SkX. 
The nonzero $\chi^{\rm sc}$ is clearly seen in the averaged scalar chirality in Fig.~\ref{Fig:3QI}(c), where $\chi_{\bm{R}}$ is distributed in a checkerboard way. 
Indeed, the averaged chirality structure factor is characterized by the dominant peaks at $\bm{Q}_1$ and $\bm{Q}_3$ and the subdominant peak at $\bm{Q}_2$. 
One finds that there is an imbalance between the regions with the positive and negative chiralities in a magnetic field. 
The similar argument holds for the AFM staking in Fig.~\ref{Fig:3QI}(f). 
We note that such a uniform chirality is not obtained in the single-layer case, $J_{\parallel}=0$. 
Thus, the emergent $\chi^{\rm sc}$ might be due to the layered structure with the different ordering vectors, which is brought about by a subtle balance between the staggered DM and interlayer interactions. 

The 3$Q'$ I state turns into the 1$Q$ state with jumps of $M^z_{\eta}$ and $\chi_{\eta}^{\rm sc}$ by increasing $|J_{\parallel}|$, whose critical values are similar to both FM and AFM interactions. 
In almost all the regions, the 3$Q'$ I state changes into the SkX upon increasing $H$. 
Thus, the appearance of the 3$Q'$ I phase below the SkX indicates the importance of the bilayer nature, which have not been found in the single-layer case ($J_{\parallel}=0$). 
Moreover, the different sign of the topological Hall signal can be observed in experiments owing to the different sign of $\chi^{\rm sc}$.

\subsection{High-field triple-$Q$ states}
\label{sec:High-field triple-$Q$ states}

\begin{figure*}[htb!]
\begin{center}
\includegraphics[width=1.0 \hsize]{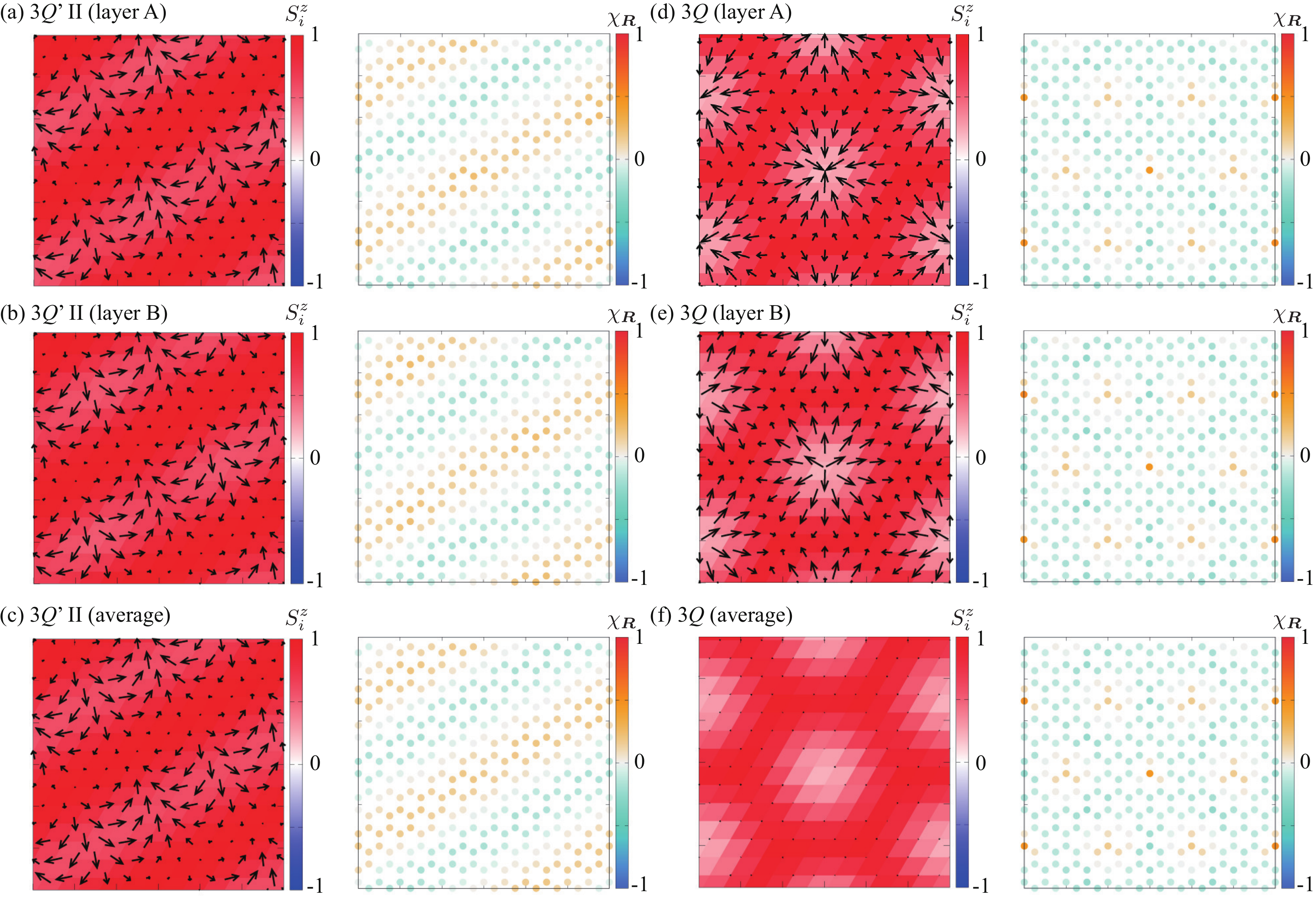} 
\caption{
\label{Fig:highfield}
(Left) Real-space spin configurations on (a), (d) the layer A and (b), (e) the layer B in (a), (b) the 3$Q'$ II state at $J_{\parallel}=-1$ and $H=1.6$ and (d), (e) the 3$Q$ state at $J_{\parallel}=0.5$ and $H=2.1$. 
(c), (f) The averaged spin configurations for the layers A and B in (c) the the 3$Q'$ II state and (f) the 3$Q$ state. 
The arrows represent the $xy$ components of the spin moment and the color shows the $z$ component. 
(Right) Real-space scalar chirality configurations calculated at each triangle plaquette. 
}
\end{center}
\end{figure*}

We discuss the magnetic phases in the high-field region, which are obtained by increasing the magnetic field from the SkX phase. 
In contrast to the SkX in Sec.~\ref{sec:Skyrmion crystal} and the 3$Q'$ I state in Sec.~\ref{sec:Low-field triple-$Q$ state}, the high-field phases are different for the FM and AFM interactions, as shown in Fig.~\ref{Fig:PD}. 
In the case of the FM stacking, the 3$Q'$ II state is stabilized between the SkX and the fully-polarized state. 
In this state, the spin configuration is characterized by the dominant double-$Q$ peaks at $\bm{Q}_1$ and $\bm{Q}_3$ and the subdominant peak at $\bm{Q}_2$ in both layers A and B, as shown in Fig.~\ref{Fig:SkX_sq}(e). 
The 3$Q'$ II state is accompanied by the scalar chirality density waves with the dominant $\bm{Q}_2$ component, whose magnitude is much larger than that of the subdominant $\bm{Q}_1$ and $\bm{Q}_3$ components. 
Thus, there is a stripe modulation of the scalar chirality in real space, as shown in Figs.~\ref{Fig:highfield}(a) and \ref{Fig:highfield}(b). 
Since the same spin and chirality configurations are realized in both layers, as shown in Figs.~\ref{Fig:highfield}(a) and \ref{Fig:highfield}(b), no intriguing averaged spin and chirality configurations appear in Fig.~\ref{Fig:highfield}(c). 

For the AFM staking, the 3$Q$ state is stabilized between the SkX and the fully-polarized state. 
The 3$Q$ state shows the triple-$Q$ peaks in both $xy$ and $z$ components of the spin structure factor with equal intensity. 
The real-space spin configurations on the layers A and B are shown in Figs.~\ref{Fig:highfield}(d) and \ref{Fig:highfield}(e), both of which consist of the triangular lattice of the vortices with $S_i^z >0$.   
This 3$Q$ state exhibits nonzero scalar chirality $\chi_\eta^{\rm sc}$ in each layer. 
As the directions of inplane spins are opposite for the layers A and B, the $xy$ spin component is canceled out in the averaged spin configuration, as shown in Fig.~\ref{Fig:highfield}(f). 
The resultant $z$ spin and chirality configurations resemble those in the AFM-coupled SkX in Fig.~\ref{Fig:SkX}(f).

\begin{figure*}[htb!]
\begin{center}
\includegraphics[width=1.0 \hsize]{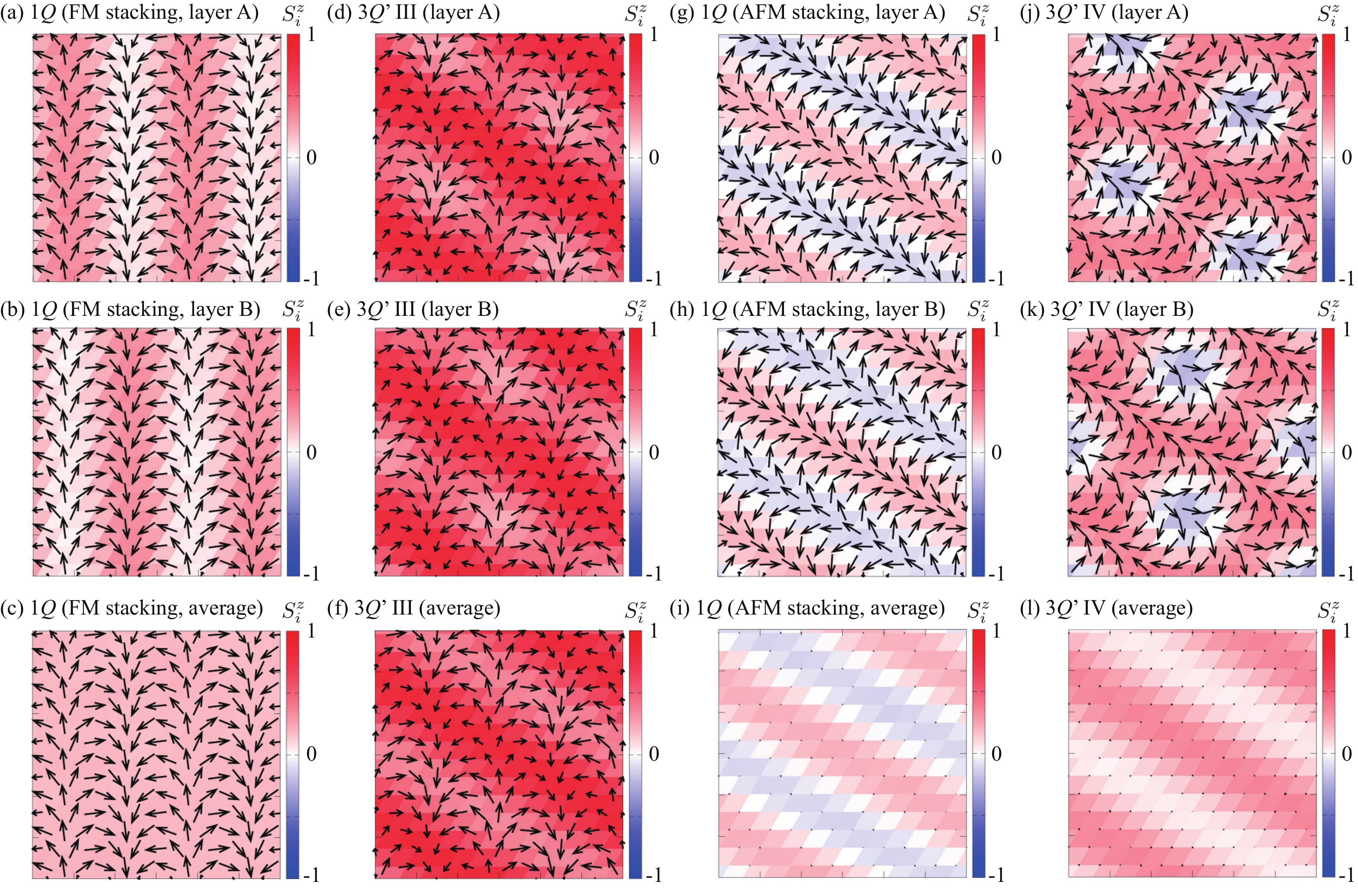} 
\caption{
\label{Fig:Other_spin}
(Left) Real-space spin configurations on (a), (d), (g), (j) the layer A and (b), (e), (h), (k) the layer B in (a), (b) the 1$Q$ state at $J_{\parallel}=-1$ and $H=0.4$, (d), (e) the 3$Q'$ III state at $J_{\parallel}=-1$ and $H=1.2$, (g), (h) the 1$Q$ state at $J_{\parallel}=1$ and $H=0.2$, and (j), (k) the 3$Q'$ IV state at $J_{\parallel}=1$ and $H=0.8$. 
(c), (f), (i), (l) The averaged spin configurations for the layers A and B in (c) the the 1$Q$  state, (f) the 3$Q'$ III state, (i) the 1$Q$ state, and (l) the 3$Q'$ IV state. 
The arrows represent the $xy$ components of the spin moment and the color shows the $z$ component. 
}
\end{center}
\end{figure*}

\section{Summary}
\label{sec:Summary}

To summarize, we have investigated the multiple-$Q$ magnetism in the centrosymmetric bilayer structure. 
We focused on the layer degree of freedom with the layer-dependent staggered DM interaction that arises from the absence of local inversion symmetry. 
By performing the simulated annealing for the spin model on the bilayer triangular lattice, we found that the SkXs are stabilized for both FM and AFM interlayer couplings. 
The obtained SkXs in the bilayer system consist of the SkXs with different helicities in each layer. 
Although the real-space spin and scalar chirality configurations in the SkXs are different for the FM and AFM cases, both the SkXs are characterized by the quantized skyrmion number. 
We also found various multiple-$Q$ states by changing the layer-dependent DM interaction and the interlayer exchange coupling. 
In particular, we showed that the low-field spiral states realized in the single layer are modulated so as to have the multiple-$Q$ spin components and the scalar chirality by the interplay between the staggered DM and the interlayer exchange interactions. 

The present result indicates that the layer degree of freedom can be a source of inducing the SkX and the other multiple-$Q$ spin states. 
The key ingredient is the layer-dependent DM interaction, which exists even in the centrosymmetric systems. 
Such a site-dependent DM interaction is found in not only the layered system but also the bulk system with the sublattice degree of freedom, such as the honeycomb and kagome structures. 
In addition, the centrosymmetric systems where the magnetic ions are located at the Wyckoff position without spatial inversion symmetry, such as $2e$ site in the space group $P6/mmm$ (\#191), are promising.
Thus, further intriguing multiple-$Q$ orderings including the SkX are expected by taking into account the layer/sublattice-dependent DM interaction.

\appendix
\section{Spin configurations and structure factors of the other phases}
\label{sec:Spin configurations and structure factors of the other phases}

\begin{figure*}[htb!]
\begin{center}
\includegraphics[width=1.0 \hsize]{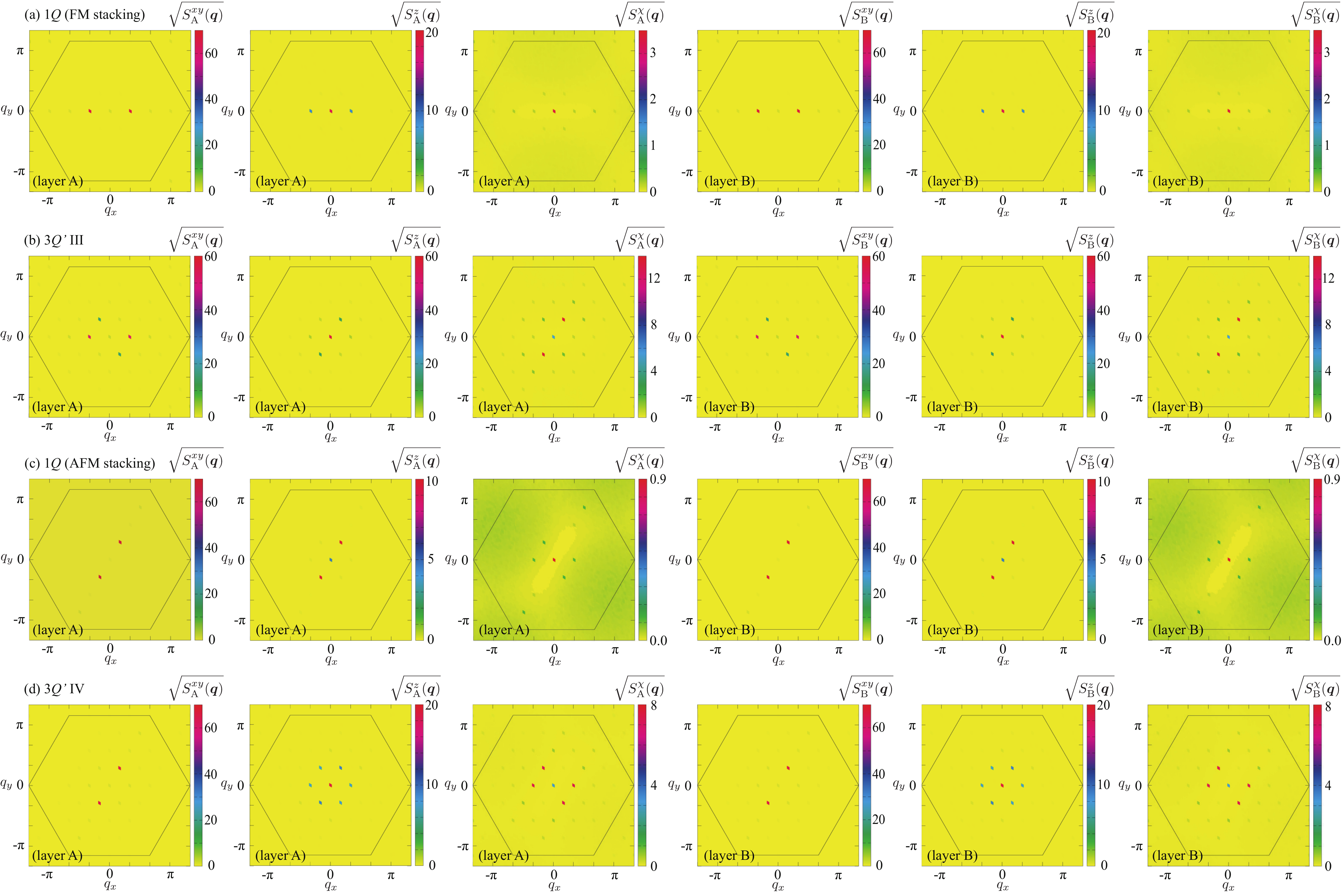} 
\caption{
\label{Fig:Other_sq}
(Left and second left) The square root of the $xy$ and $z$ components of the spin structure factor, respectively, in (a) the 1$Q$ state at $J_{\parallel}=-1$ and $H=0.4$, (b) the 3$Q'$ III state at $J_{\parallel}=-1$ and $H=1.2$, (c) the 1$Q$ state at $J_{\parallel}=1$ and $H=0.2$, and (d) the 3$Q'$ IV state at $J_{\parallel}=1$ and $H=0.8$ for the layer A.
Black hexagons represent the first Brillouin zone.   
(Middle left) The square root of the chirality structure factors for the layerA. 
The right three panels represent the data for the layer B corresponding to the left three ones. 
}
\end{center}
\end{figure*}

In this Appendix, we show the spin configurations in the 1$Q$ state, the 3$Q'$ III state, and the 3$Q'$ IV state, which are stabilized only for large $|J_{\parallel}|$.
Figures~\ref{Fig:Other_spin} show the real-space spin configurations in (a), (b) the 1$Q$ state at $J_{\parallel}=-1$ and $H=0.4$, (d), (e) the 3$Q'$ III state at $J_{\parallel}=-1$ and $H=1.2$, (g), (h) the 1$Q$ state at $J_{\parallel}=1$ and $H=0.2$, and (j), (k) the 3$Q'$ IV state at $J_{\parallel}=1$ and $H=0.8$, which are obtained by simulated annealing. 
We also show the averaged spin configurations over the layers in Figs.~\ref{Fig:Other_spin}(c), \ref{Fig:Other_spin}(f), \ref{Fig:Other_spin}(i), and \ref{Fig:Other_spin}(l) in each state. 

The spin and chirality structure factors corresponding to Fig.~\ref{Fig:Other_spin} are shown in Fig.~\ref{Fig:Other_sq}. 
The spin configurations in the 1$Q$ state in Figs.~\ref{Fig:Other_sq}(a) and \ref{Fig:Other_sq}(c) are characterized by the single-$Q$ peak in the spin structure factor. 
The nonzero $S^{\chi}_\eta(\bm{0})$ indicates the staggered chirality configuration on upward and downward triangles on the triangular lattice. 
The 3$Q'$ III state in Fig.~\ref{Fig:Other_sq}(b) and the 3$Q'$ IV state in Fig.~\ref{Fig:Other_sq}(d) show the triple-$Q$ peak structures with different intensities in both $xy$ and $z$ spin components. 
Both states show the chirality density waves as shown in $S^{\chi}_\eta(\bm{q})$. 

\section{Results for different $D$}
\label{sec:Results for different D}

\begin{figure}[htb!]
\begin{center}
\includegraphics[width=1.0 \hsize]{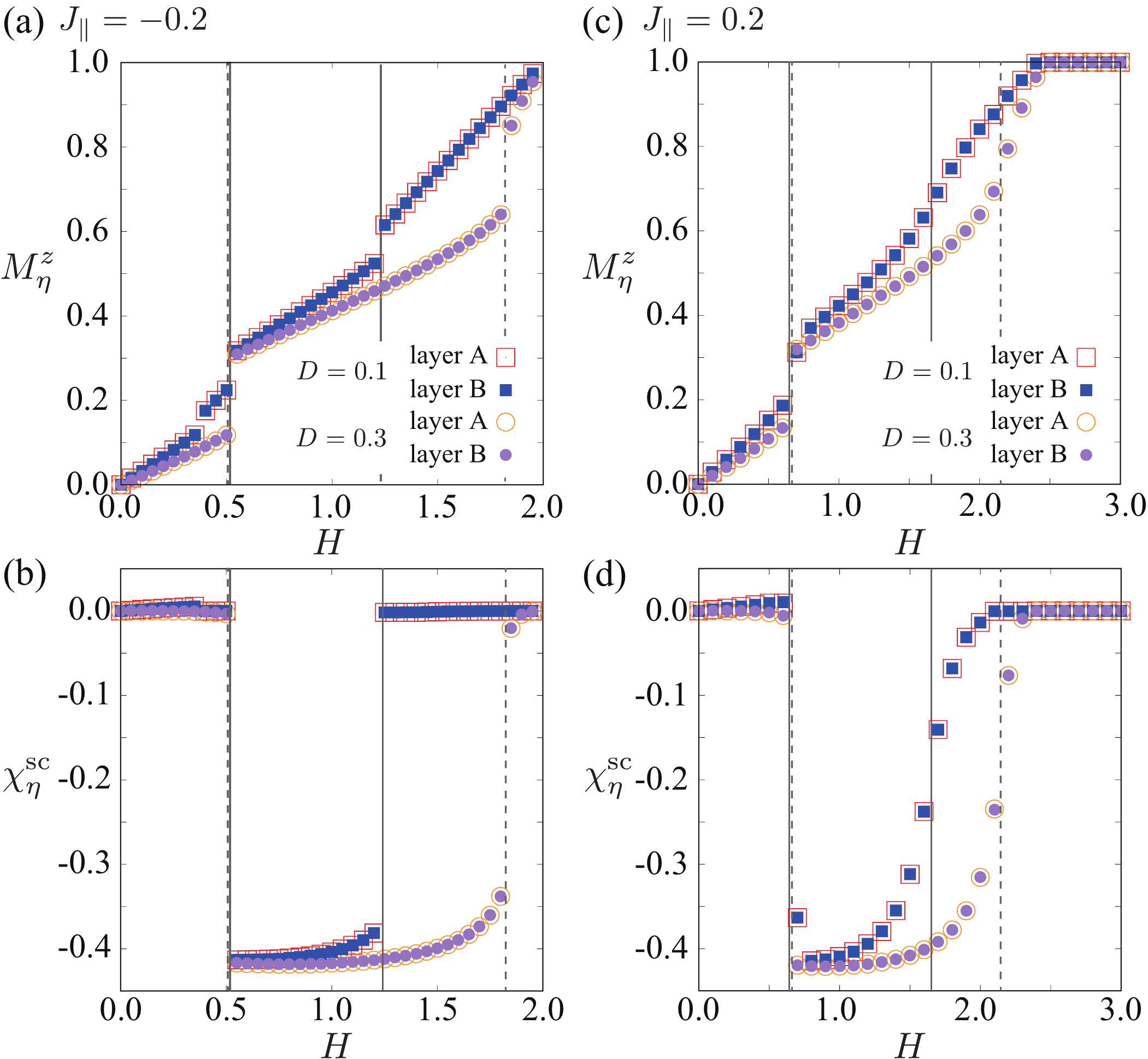} 
\caption{
\label{Fig:Mag_chirality_otherDM}
$H$ dependence of (a), (c) the magnetization $M^z_\eta$ and (b), (d) the scalar chirality $\chi^{\rm sc}_\eta$ for the layers $\eta=$A and B at (a), (b) $J_{\parallel}=-0.2$ and (c), (d) $J_{\parallel}=0.2$ for $D=0.1$ and $0.3$. 
The solid (dashed) vertical lines represent the phase transitions between the SkX and the other magnetic states for the small (large) values of $D$. 
}
\end{center}
\end{figure}

We show the results while changing the magnitude of the DM interaction in the model Hamiltonian $\tilde{\mathcal{H}}$. 
Figures~\ref{Fig:Mag_chirality_otherDM}(a)-(d) show the $H$ dependence of $M^z_\eta$ [(a) and (c)] and $\chi^{\rm sc}_\eta$ [(b) and (d)] for $\eta=$A and B for $D=0.1$ and $0.3$. 
The results for the FM stacking and the AFM stacking are shown in Figs.~\ref{Fig:Mag_chirality_otherDM}(a),(b) and Figs.~\ref{Fig:Mag_chirality_otherDM}(c),(d), respectively. 
The overall behavior against $H$ is similar to that in Fig.~\ref{Fig:Mag_chirality}. 
Meanwhile, one finds that the large DM interaction enhances the stability of the SkX in both cases, which is common to the case in noncentrosymmetric single-layer systems.

\begin{acknowledgments}
This research was supported by JSPS KAKENHI Grants Numbers JP19K03752, JP19H01834, JP21H01037, and by JST PRESTO (JPMJPR20L8). 
Parts of the numerical calculations were performed in the supercomputing systems in ISSP, the University of Tokyo.
\end{acknowledgments}

\bibliographystyle{apsrev}
\bibliography{ref.bib}

\begin{thebibliography}{128}
\bibitem[{com()}]{comment_sixfold}
\bibinfo{note}{It is noted that sixfold rotational symmetry is slightly broken
  in the AFM-coupled SkX, since the skyrmion core lies on the center of the
  triangle plaquette~\cite{hayami2021locking}.}
\end{thebibliography}

\end{document}